# FPGA-based Multi-Chip Module for High-Performance Computing

Yann Beilliard[1,2], Maxime Godard[1,2], Aggelos Ioannou[4], Astrinos Damianakis[4], Michael Ligerakis[4], Iakovos Mavroidis[4], Pierre-Yves Martinez[3], David Danovitch[1,2], Julien Sylvestre[1,2], Dominique Drouin[1,2]

[1]*Institut Interdisciplinaire d'Innovation Technologique (3IT), Université de Sherbrooke, Canada*
[2]*Laboratoire Nanotechnologies Nanosystemes (LN2) – CNRS, Canada*
[3]*CEA-LETI, 38054 Grenoble, France*
[4]*Foundation for Research and Technology (FORTH) – Hellas, Heraklion, Greece*
Corresponding author: yann.beilliard@usherbrooke.ca

**Abstract** – Current integration, architectural design and manufacturing technologies are not suited for the computing density and power efficiency requested by Exascale computing. New approaches in hardware architecture are thus needed to overcome the technological barriers preventing the transition to the Exascale era. This article presents the design, fabrication and characterization of a first ExaNoDe multi-chip module prototype, aiming to enable Exascale computing through highly integrated heterogeneous package.

## I. Introduction

The most powerful supercomputers built so far, ranked in the *Top500 Supercomputers list* [1], exhibit peak floating point (FP) performance in the range of $10^{15}$, corresponding to Petascale computing. Even such performance will soon not meet the needs of increasingly complex scientific problems, such as computational biology, climate change or energy. The scientific community is thus already investigating the development of supercomputers capable to reach Exascale performance ($10^{18}$ FP operations per second). This is a worldwide trend, with programs in USA, China, Japan and Europe. The European H2020 ExaNoDe research project [2] investigates, develops, integrates and validates the building blocks for a compute element that enables Exascale performance. One of the main principles of the ExaNoDe project is to combine advanced packaging technologies with three dimensional (3D) integration of high-performance dies [3] to deliver a prototype-level system demonstrating that those technologies are promising candidates towards the definition of an Exascale compute node. The prototype envisioned is an advanced multi-chip module (MCM) featuring compute FPGAs and an active silicon interposer on which chiplets are stacked, as illustrated in Fig 1*a*. The chiplet approach combined with the silicon interposer technology will lead to a higher level of integration of multiple compute units and memories within the same compute node. This high-density heterogenous integration will enlarge the application scope of a single compute node, while scaling-up the performance of a compute system towards Exascale performance. Before achieving successful assembly of this fully-fledged MCM prototype called Mk II, a preliminary version Mk I composed of 2 Xilinx Zynq Ultrascale+ MPSoC (Fig 1*b*) was first fabricated to validate key building blocks through morphological and electrical investigations. This paper presents the successful fabrication of MCM Mk I prototypes. We will first describe the design and assembly process of the MCMs, then morphological investigations of the chip join quality will be presented. Finally, electrical tests results will be discussed.

## II. Multi-Chip Module Development

### a. Laminate Substrate Design

The MCM Mk I being essentially a short loop version of the MCM Mk II, both prototypes share the same laminate substrate. The latter was thus design to efficiently interconnect together both FPGAs, the silicon interposer as well as their corresponding decoupling capacitors. For integration density reasons, the choice was made to use a 5-2-5 laminate substrate with a non-standard 68.5 × 55 mm² area. FPGA #1 and #2 are located at the north end of the substrate with respect to the interposer. As visible in Fig 2, the FPGA #2 was rotated 90 ° clock wise compared to FPGA #1 in order to greatly optimize the routing of low-voltage differential signals (LVDS) and allowing the connection of the FPGA #2 to its small outline dual in-line memory module (SODIMM) located on the board. A significant number of LVDS and chip-to-chip nets were assigned accordingly to the corresponding and available banks of the FPGA for straightforward high-density routing.

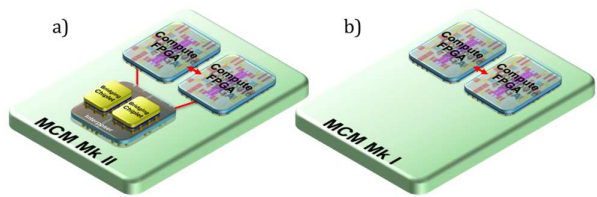

*Fig 1. Schematic representations of a) the full ExaNoDe's MCM Mk II prototype composed of 2 FPGAs, 1 silicon interposer on which 2 chiplets are stacked and b) the preliminary MCM Mk I.*

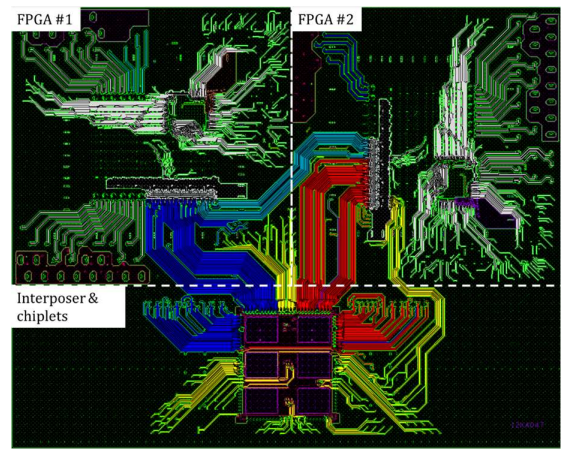

*Fig 2. Top-view of one of the ten routing layers inside the laminate substrate, connecting all components of the MCM.*

### b. Assembly Process

After fabrication of the laminate substrates by Kyocera, the first step of the MCM Mk I assembly process consisted in the soldering of both FPGAs and their decoupling capacitors. A flux material was thus dispensed on the bumps of the laminate, promoting defect free soldering of the C4 connections and good placement of the components. A mass reflow was then performed to permanently solder the components. Afterwards, assisted by capillary forces, an underfill (CUF) material was dispensed under the FPGAs to encapsulate the C4 bumps at the chip/laminate interface (Fig 5). The purpose of the CUF is to enhance the interconnection strength and compensate for differences in thermal expansion coefficient that could lead to cracks and delamination. The 1 mm pitch ball grid array (BGA) was then fabricated on the backside of the laminate with 600 µm wide solder balls. Finally, thermal interface material was dispensed on each FPGA before the capping of the MCMs with a 1 mm thick Cu/Ni lid.

## III. Characterizations

### a. Morphological investigations

Multiple morphological investigations were carried out to validate the assembly quality of the modules. Some of the results are shown in Fig 3. Scanning acoustic microscopy and x-ray inspections (Fig 3*a*) were conducted on the FPGAs to control the CUF dispensing and the soldering of the C4 connections. No voids, non-wet or short circuit were found. Cross-sections of the FPGAs were also performed without finding any voids or cracks in the well-aligned interconnections as visible in Fig 3*b*. Visual inspections of the BGA revealed well-formed solder balls without short-circuit after reflow (Fig 3*c*). Finally, Therm-Moiré measurements on assembled MCM Mk I indicated that the maximum warpage was contained to a low value of 131 µm (Fig 3*d*), which is below the 220 µm recommended by JEITA [4] for card attach with a 1 µm pitch BGA.

### b. Electrical tests

The assembled MCMs Mk I were mounted on daughter boards (DB) by FORTH with a standard reflow process. As shown in Fig 5, each DB was plugged into a small carrier board called Minifeeder, which provides 10 high-speed small form-factor pluggable (SFP+) transceivers, 1 GigE interface, universal asynchronous receiver transmitters (UARTs) and several general purpose I/Os (GPIOs), which are directly connected to the board. The DDR memories on the 4 SODIMMs of the DB were successfully tested by running extensive Xilinx memory tests with clock frequencies of 1866 MHz and 2133 MHz. The JTAG chain connectivity was functional and used to program the FPGAs. All 4 FPGAs of the 2 MCMs Mk I were thus programmed with the Xilinx integrated bit error ratio test (IBERT) tailored for this board for links testing. All intra-board high-speed links between all 4 FPGAs were stable at 10 Gbps, even under the more demanding 31-bit PRBS (Pseudorandom Binary Sequence) tests.

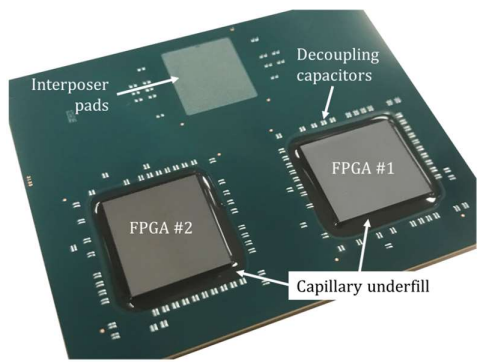

*Fig 5. Photograph of an assembled MCM Mk I, composed of 2 Xilinx Zynq Ultrascale+ MPSoC and their decoupling capacitors.*

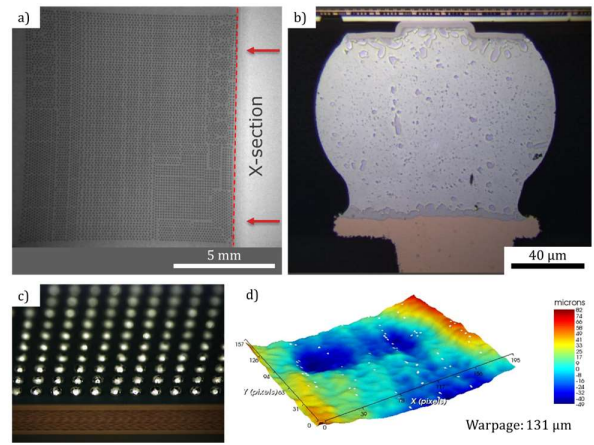

*Fig 3. a) Top-down x-ray image of a soldered FPGA showing no short circuit. b) Cross-section micrograph of a C4 connection indicating excellent alignment and no void or crack. c) Tilted view of a defect-free BGA. d) Warpage inspection of an assembled MCM Mk I by Therm-Moiré.*

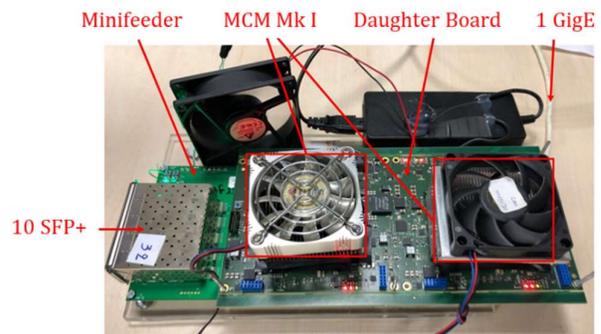

*Fig 4. Photograph of a fully assembled daughter board used for testing the performance and functionality of 2 MCMs Mk I.*

## IV. Conclusion

Reaching Exascale computing performance demands to develop energy-efficient high-density architectures. To that end, we have demonstrated successful assembly and card attach of ExaNoDe FPGA-based MCM preliminary prototypes dedicated to Exascale computing applications. The next step of this project will consist in the demonstration of MCM prototypes combining advanced packaging and 3D integration, featuring 2 Xilinx Zynq Ultrascale+ MPSoCs, an active silicon interposer on which 2 chiplets will be stacked.

### Acknowledgement

This work was supported by the ExaNoDe project that has received funding from the European Union's Horizon 2020 research and innovation programme under grant agreement No. 671578.

### Reference

[1] TOP500 Supercomputers List: https://www.top500.org.
[2] H2020 ExaNoDe website: http://exanode.eu.
[3] A. Rigo *et al.*, *ECDSD*, 2017, pp. 486–493.
[4] ED-7306: home.jeita.or.jp/tsc/std-pdf/ED-7306_E.pdf